\documentclass[
reprint,
amsmath,amssymb,
aps,
prb,
]{revtex4-2}

\usepackage{graphicx}
\usepackage{dcolumn}
\usepackage{bm}
\usepackage{mhchem} 
\usepackage{booktabs} 

\begin{document}

\title{An Exchange-Correlation Functional for Fast and Accurate Modeling of Ferroelectric Perovskites}

\author{Owain T. Beynon}

\author{Chiara Gattinoni}%
 \email{chiara.gattinoni@kcl.ac.uk}
\affiliation{Theory and Simulation of Condensed Matter Group, Department of Physics, King's College London, London, UK, WC2R 2LS}

\begin{abstract}

We present a novel exchange correlation functional, C09x-PBEc, which combines C09 exchange with PBE correlation, to accurately model the ferroelectric properties of perovskites while retaining the computational efficiency of GGA functionals. 
With a growing interest in developing machine learning interatomic potentials (MLIPs) to model large-scale ferroelectric systems of technological relevance, it is important to scrutinise the density functional theory exchange-correlation functionals which are used to compute the forces, energies and stresses the MLIP is trained on.
Using the example of the prototypical ferroelectrics lead titanate, \ce{PbTiO3}, and barium titanate, \ce{BaTiO3} we show that many widely used functionals tend to overestimate their lattice constants and spontaneous polarization.
Conversely, non-local van der Waals functionals with C09 exchange accurately capture these properties compared to experiment, but with a larger computational overhead than, for example, GGA.
We show that C09x-PBEc combines the accuracy provided by the C09 exchange with the computational affordability of GGA, making it an excellent candidate to be used in the training of MLIPs for ferroelectric perovskites.
We also demonstrate that an MLIP trained using C09x-PBEc accurately reproduces the ferroelectric-to-paraelectric phase transition temperature of \ce{PbTiO3} with respect to experiment, showing a marked improvement on MLIPs trained using GGA.

\end{abstract}


\maketitle

\section{Introduction}


The switchable and tunable spontaneous polarization exhibited by ferroelectric perovskites makes them suitable for a wide range of applications in electronics as capacitors, thermistors, optoelectronic devices, non-volatile memories and piezoelectric sensors~\cite{nuraje_perovskite_2013, pacchioni_ferroelectricity_2011}.
Other emerging applications are in catalysis, superconductors, and electrochemical energy conversion~\cite{wang_perovskite-based_2024, kakekhani_ferroelectric-based_2015, xie_improving_2019, cui_surface_2016}.
Understanding the behavior of ferroelectric materials is vital to progress their use in applications; however, accurately capturing their ferroelectric properties through atomistic materials modeling is not trivial. 

Density functional theory (DFT) has been widely employed to model ferroelectric perovskites. \cite{cardonaquintero_applicability_2025, yuk_towards_2017, bilc_hybrid_2008, Wu_Cohen_PhysRevB.73.235116}. 
However, most exchange-correlation (XC) functionals tend to largely overestimate the value of the spontaneous polarization.
%
%
%
Moreover, for larger systems DFT is not suitable due to its intrinsic computational cost, and computationally less demanding force fields for molecular dynamics  investigations of ferroelectric materials have been developed~\cite{zhang_developing_2022, shin_development_2005, qi_atomistic_2016} including machine learning 
interatomic potentials (MLIPs) using, for example, 
the gaussian approximation potentials method~\cite{gigli_thermodynamics_2022} and atomic clustering expansion~\cite{sehrawat_machine-learning_2026}. 

MLIPs have emerged as a powerful implementation of artificial intelligence, \cite{elena_machine_2025, ko_recent_2023, sehrawat_machine-learning_2026} where through training models on data from $ab$ $initio$ calculations \textit{i.e.} DFT, they can simulate materials at the trained DFT level of accuracy with a significant reduction in computational overhead. 
%
%
The performance of MLIPs thus heavily depends on the XC functional employed to build the training set, as ultimately the accuracy of the model to describe the potential energy surface is restricted to that of DFT.
Considering that training requires many thousands of structures from DFT, XC functionals which are computationally inexpensive are preferred. 
On the other hand, material properties required for the problem at hand need to be accurately reproduced, and thus the choice of XC functional must be subject to careful consideration. 
%

Of the XC functionals used in DFT investigations of ferroelectric perovskites, varying levels of theory capture ferroelectric properties of the material with varying accuracy. 
Indeed, the majority of XC functionals, for example LDA, GGA and meta-GGAs, including SCAN, overestimate the value of the ferroelectric polarization in \ce{PbTiO3} and \ce{BaTiO3} by as much as 25\% \cite{zhang_comparative_2017}. 
%
%
Employing hybrid functionals produces a more accurate representation of the structural properties, polarization, and band gap of ferroelectric materials but at a huge computational overhead \cite{bilc_hybrid_2008}. 
%
Functionals with non-local vdW-DF and vdW-DF2 correlation, which account for long range dispersion interactions, improve accuracy in the reproduction of the experimental polarization whilst avoiding the computational expense of hybrid functionals \cite{cardonaquintero_applicability_2025, yuk_putting_2024, schroder_vdw-df_2017, berland_van_2014}, especially when coupled to Cooper's exchange term (C09) \cite{cooper_van_2010}.
In the latter the prediction of the polarization is within 4\% of the experimental value for \ce{PbTiO3} and 11\% for \ce{BaTiO3} ~\cite{yuk_towards_2017}. 

\begin{figure}
    \centering
    \includegraphics[width=1.0\linewidth]{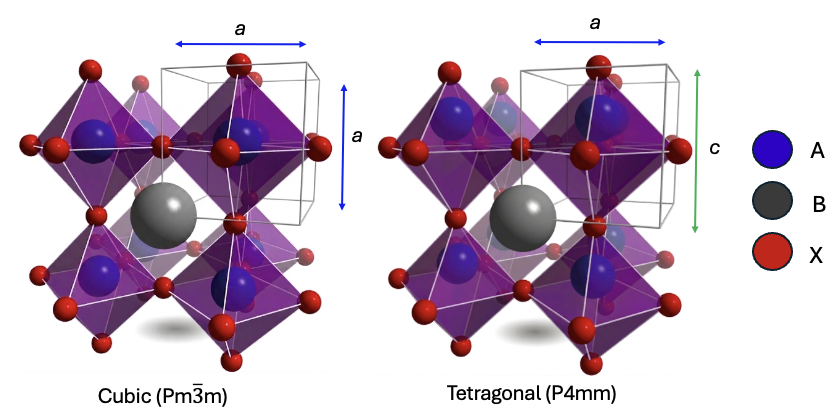} 
    \caption{Structures of tetragonal and cubic \ce{PbTiO3} and \ce{BaTiO3}. Red, blue, and grey spheres represent O, Ti, and Pb/Ba atoms, respectively.}
    \label{Figure1}
\end{figure}

Herein, we examine a range of XC functionals and their accuracy on modeling the properties of two perovskite structures, \ce{PbTiO3} and \ce{BaTiO3} (Fig. 1). 
We find that the C09 exchange is crucial in capturing accurate lattice parameters and spontaneous polarization value. Thus, we present a new functional that includes C09 exchange to standard PBE correlation, which we have termed C09x-PBEc,
which provides an excellent trade off between robustness and computational efficiency, making it ideal to be used in the training of MLIPs.

\section{Computational Methods}

DFT calculations were performed with the plane-wave Quantum Espresso software package~\cite{giannozzi_quantum_2009}. 
The energy cut-off for the plane-wave basis set was 50 Ry, 500 Ry for the charge density cut-off, and ultrasoft pseudopotentials were used to replace core electrons~\cite{vanderbilt_soft_1990}. 
Calculations were performed with a converged Monkhorst-Pack $k$-point sampling grid of $4\times4\times4$~\cite{monkhorst_special_1976} for the 5-atom unit cell. 
For a comparative investigation of the effect of exchange and correlation on the ferroelectric properties, a range of XC functionals were benchmarked and evaluated. 
At the general gradient approximation level of theory,  the XC functionals of Perdew-Burke-Ernzerhof (PBE)~ \cite{perdew_generalized_1996}, the version reparametrized for solids (PBEsol)~\cite{perdew_restoring_2008}, and the revised version (revPBE)~\cite{zhang_comment_1998} were used.
Furthermore, functionals with non-local van der Waals correlation were considered: vdW-DF~\cite{ dion_van_2004} (with revPBE exchange) and vdW-DF2~\cite{, PhysRevB.82.081101} (with revised PW86 exchange). 
Finally, the non van der Waals correlation with the C09 exchange term proposed by Cooper et al.~\cite{ cooper_van_2010} were used (vdW-DF-C09 and vdW-DF2-C09).

The Berry phase approach~\cite{spaldin_beginners_2012} was used to calculate the spontaneous polarization of bulk structures. 
The Born effective charge tensors (\(Z_{\alpha \beta }^{*}\)) were calculated at the zone center (\(\Gamma \)-point) by solving the self-consistent linear response equations to a homogeneous electric field. \

Unit cells for cubic and tetragonal \ce{PbTiO3} and \ce{BaTiO3} were obtained from the Materials Project database~\cite{horton_accelerated_2025}, 
and models were built and manipulated using the Atomic Simulation Environment (ASE)~\cite{hjorth_larsen_atomic_2017}.
All structures were subject to full unit cell optimizations performed using the Broyden-Fletcher-Goldfarb-Shanno (BFGS) algorithm with convergence reached when the forces on all atoms are less than a criteria of 0.01 $\mathrm{eV\,\AA^{-1}}$~\cite{broyden_convergence_1970, fletcher_new_1970, goldfarb_family_1970, shanno_conditioning_1970}.

An MLIPs was trained for \ce{PbTiO3} using the MACE architecture~\cite{batatia_mace_2022}. 
A training set of  551 structures was obtained via molecular dynamics (MD) simulations on unit cells and $3\times3\times3$ supercells of the cubic and tetragonal structures of \ce{PbTiO3} using the MACE foundational model with and without dispersion corrections, \textit{MACE-MPA-0-D3}. \cite{batatia_foundation_2025}
%
%
The potential energy surface was sampled using NPT and NVT ensembles on temperatures of 300, 400, 500, 600, 700, 800 and 900 K, using Langevin dynamics for NVT, Bernedsen dynamics for NPT, and a timestep of 1 fs for both ensembles. 
Single point energy calculations were then performed using structures obtained from the MD using Quantum Espresso and the C09x-PBEc functional.

For the proposed C09x-PBEc, the full exchange-correlation functional takes the form of: 
\begin{equation}
E_{xc}^{\mathrm{C09x-PBEc}}
=
E_{x}^{\mathrm{C09}}
+
E_{c}^{\mathrm{PBE}}
\end{equation}
where, for the C09 exchange term, $E_{x}^{\mathrm{C09}}$,  the total exchange energy is defined by modifying the local density approximation (LDA) via a gradient-dependent exchange enhancement factor, $F_x(s)$, as defined in Ref.~\cite{cooper_van_2010}:

\begin{equation}
E_x^{\text{C09}} = \int d^3r \, \rho(\mathbf{r}) \epsilon_x^{\text{unif}}(\rho) F_x(s)
\end{equation}

Here, $\epsilon_x^{\text{unif}}(\rho) = -\frac{3}{4\pi}(3\pi^2\rho)^{1/3}$ represents the conventional Slater exchange energy density for a uniform electron gas, and $s = |\nabla\rho| / [2(3\pi^2)^{1/3}\rho^{4/3}]$ is the dimensionless reduced density gradient, and $F_x(s)$, is defined as: 

\begin{equation}
F_x(s) = 1 + \mu s^2 \, e^{-\alpha s^2} + \kappa \left( 1 - e^{-\alpha s^2 /2} \right)
\end{equation}

where $\kappa$ is 1.245, $\mu$ is 0.0617, and $\alpha$ is 0.0483~\cite{cooper_van_2010}. 
$E_{c}^{\mathrm{PBE}}$, is the standard PBE correction to the LDA correlation energy \cite{perdew_generalized_1996}.  

\section{Results and Discussion}

\begin{table*}[ht]
\caption{Calculated structural properties ($a$ and $c$ lattice parameters) of cubic and tetragonal \ce{PbTiO3} and \ce{BaTiO3} compared to experiment.}
\centering
\label{tab:structural_split}
\resizebox{\textwidth}{!}{%
\begin{tabular}{llcccccccccc}
\toprule
\textbf{Phase} & \textbf{Properties} & \textbf{PBE} & \textbf{PBEsol} & \textbf{revPBE} & \textbf{vdW-DF} & \textbf{vdW-DF2} & \textbf{vdW-DF-C09} & \textbf{vdW-DF2-C09} & \textbf{C09x-PBEc} & \textbf{Exp.} \\
\midrule
\multicolumn{11}{c}{\textbf{PbTiO$_3$}} \\
\midrule
Cubic       & a / \AA         & 3.97 & 3.92 & 3.99 & 4.00 & 4.03 & 3.92 & 3.93 & \textbf{3.90} & 3.93 \cite{https://doi.org/10.1107/S0021889879011754} \\
Tetragonal  & a / \AA         & 3.84 & 3.87 & 3.86 & 3.88 & 3.86 & 3.89 & 3.89 & \textbf{3.87} & 3.90 \cite{PhysRev.97.1179} \\
            & c / \AA         & 4.77 & 4.22 & 4.95 & 4.86 & 4.91 & 4.08 & 4.14 & \textbf{4.07} & 4.15 \cite{PhysRev.97.1179} \\
\midrule
\multicolumn{11}{c}{\textbf{BaTiO$_3$}} \\
\midrule
Cubic       & a / \AA         & 4.02 & 3.98 & 4.05 & 4.07 & 4.09 & 3.98 & 3.99 & \textbf{3.95} & 4.00 \cite{Hellwege1969Ferroelectrics} \\
Tetragonal  & a / \AA         & 3.98 & 3.96 & 3.99 & 4.02 & 4.03 & 3.96 & 3.97 & \textbf{3.94} & 3.99 \cite{doi:10.1021/j100112a043} \\
            & c / \AA         & 4.22 & 4.07 & 4.38 & 4.34 & 4.38 & 4.05 & 4.07 & \textbf{4.03} & 4.04 \cite{doi:10.1021/j100112a043}\\
\bottomrule
\end{tabular}
}
\end{table*}

%
%
We consider two prototypical ferroelectric perovskite systems: \ce{PbTiO3} and \ce{BaTiO3}. 
In these two materials the spontaneous polarization depends on two different mechanisms, due to the nature of the underlying chemical bonding~\cite{zhang_comparative_2017}. 
In \ce{PbTiO3}, the polarization has a strong covalent component and is driven by the hybridization between $6s$ lone-pair of the A-site Pb and the O $2p$ orbitals, leading to anomalously high Born effective charges and a large spontaneous polarization of 0.75 Cm$^{-2}$ (see Table II). 
Conversely, \ce{BaTiO3} has a smaller spontaneous polarization of 0.27 Cm$^{-2}$ and is driven by long-range ionic interactions, with hybridization mainly between empty B-site Ti $3d$ orbitals and O $2p$, where Ba-O bonds remain ionic. 
Thus by considering both \ce{PbTiO3} and \ce{BaTiO3}, we compare the performance of DFT functionals for accurately modeling two different models of polarization in ferroelectric perovskites. 

The calculated lattice parameters for both \ce{PbTiO3} and \ce{BaTiO3} and the selected range of functionals is shown in (Table \ref{tab:structural_split}).
For the cubic phase of \ce{PbTiO3} and \ce{BaTiO3}, all functional reproduce the experimental lattice parameter of 3.93 \AA\ and 4.00 \AA, within 2.54\%  and 2.25 \%, respectively. Where for both perovskites systems vdW-DF and vdW-DF2 are the least accurate, but with C09 exchange (vdW-DF/vdW-DF2-C09) the $a$ parameter is reproduced within 0.25\% and 0.5 \%. 
Similarly, the $a$ parameter in the tetragonal phase is well reproduced in both \ce{PbTiO3} and \ce{BaTiO3}, within 1.5\% (0.8\%) of the experimental value for all XC functionals.
%
Also in this case vdW-DF2-C09 provides the closest value to experiment. 
The $c$ parameter is the most sensitive to the XC functional, which is largely overestimated in both material by up to 20\% (\ce{PbTiO3} with revPBE) and 8\% (\ce{BaTiO3} with revPBE and vdW-DF2).
Conversely, vdW-DF2-C09 reproduces a $c$ parameter closest to the experimental within 0.01 \AA\ (\ce{PbTiO3}) and 0.03 \AA\ (\ce{BaTiO3}) in good agreement with previous work~\cite{yuk_towards_2017, berland_van_2014}.

\begin{table*}[ht]
\caption{Calculated spontaneous polarization ($P$) for tetragonal \ce{PbTiO3} and \ce{BaTiO3} compared to experiment.}
\centering
\label{tab:polarisation_split}
\resizebox{\textwidth}{!}{%
\begin{tabular}{llcccccccccc}
\toprule
\textbf{System} & \textbf{Properties} & \textbf{PBE} & \textbf{PBEsol} & \textbf{revPBE} & \textbf{vdW-DF} & \textbf{vdW-DF2} & \textbf{vdW-DF-C09} & \textbf{vdW-DF2-C09} & \textbf{C09x-PBEc} & \textbf{Exp.} \\
\midrule
\textbf{PbTiO$_3$} & P / Cm$^{-2}$   & 1.26 & 0.96 & 0.82 & 1.25 & 1.23 & 0.76 & 0.83 & \textbf{0.79} & 0.75 \cite{haun_thermodynamic_1987}\\
\textbf{BaTiO$_3$} & P / Cm$^{-2}$   & 0.49 & 0.38 & 0.59 & 0.54 & 0.54 & 0.34 & 0.36 & \textbf{0.33} & 0.27 \cite{PhysRev.99.1161}\\
\bottomrule
\end{tabular}
}
\end{table*}

The spontaneous polarization shows similar sensitivity to XC functionals as the $c$ lattice parameter, being largely overestimated compared with the experimental value, see Table \ref{tab:polarisation_split}.   
Indeed, for tetragonal \ce{PbTiO3}, polarization values up to 1.26 Cm$^{-2}$ (PBE) are found, and up to 0.59 (revPBE)  for \ce{BaTiO3}.
%
As with the tetragonal phase $c$ lattice parameter,  functionals with C09 exchange give the value that is the closest to experiment for both materials, with vdW-DF-C09 producing values of 0.76 Cm$^{-2}$ for \ce{PbTiO3} and 0.34 Cm$^{-2}$ for \ce{BaTiO3}, commensurate with previous studies~\cite{yuk_towards_2017}, and vdW-DF2-C09 a close second.
%
%

We have shown, as previously demonstrated~\cite{yuk_towards_2017}, that functionals with dispersion corrections in both the exchange and correlation part (vdW-DF-C09 and vdW-DF2-C09) perform best in reproducing the structural properties of \ce{PbTiO3} and \ce{BaTiO3}.
It however appears from our results that
%
the C09 exchange term is the crucial component which determines their excellent performance.
%
Indeed, vdW-DF and vdW-DF2, do not produce $c$ parameters close to experimental for \ce{PbTiO3}, but rather greatly overestimate them by $18$\% and $19$\%, respectively, performing worse than PBE (15\% error).
Also polarization values are poorly reproduced, being overestimated by over 60\% in both cases. 
Also for \ce{BaTiO3} vdW-DF and vdW-DF2 show performance comparable to that of PBE in the prediction of both the lattice parameters and the polarization value.
These results hint that dispersion corrections in the correlation alone do not provide any substantial improvement in the modeling of ferroelectric properties.


In order to confirm the greater relevance of the C09 exchange term in predicting accurate ferroelectric properties, over vdW-DF/ vdW-DF2 correlation,  we combined C09 exchange and PBE correlation, in a new functional that we call C09x-PBEc.
Results of the lattice parameters and polarization obtained optimizing the \ce{PbTiO3} and \ce{BaTiO3} unit cells with this functional are shown in the highlighted column in Table I and II.
This functional produces noticeable improvements on standard PBE and of vdW-DF/ vdW-DF2, with calculated lattice parameters within $2$\% of the experimental ones for both \ce{PbTiO3} and \ce{BaTiO3}.
%
%
The polarization $P$ is also well reproduced with C09x-PBEc, with values 0.79 Cm$^{-2}$ for \ce{PbTiO3} and 0.33 Cm$^{-2}$ for \ce{BaTiO3}  which are comparable to vdW-DF-C09 and vdW-DF2-C09, and within 0.05 Cm$^{-1}$ of the experimental value. 
%

This demonstrates that C09 exchange is the relevant component for accurate modeling, while dispersion corrections in the correlation do not produce any improvements over PBE in the calculation of structural properties.
%
%
 %
 Studies in the literature have previously examined the exchange enhancement factor (EEF), $F_x(s)$, which is a functional of the reduced density gradient (Figure S1), in PBE and C09, and its effects on modeling ferroelectric perovskites~\cite{yuk_putting_2024, cardonaquintero_applicability_2025}. 
 PBE has an increased short-range repulsion in small region of reduced density, $s$, and thus overestimates the structural properties of \ce{PbTiO3} and \ce{BaTiO3}. 
 C09 has a reduced short range repulsion in this region and at larger reduced density recovers the behavior of revPBE, leading to improved modeling of ferroelectric properties in \ce{PbTiO3} and \ce{BaTiO3} (Supporting Information (SI), Section 1).

\begin{figure}
    \centering
    \includegraphics[width=1.0\linewidth]{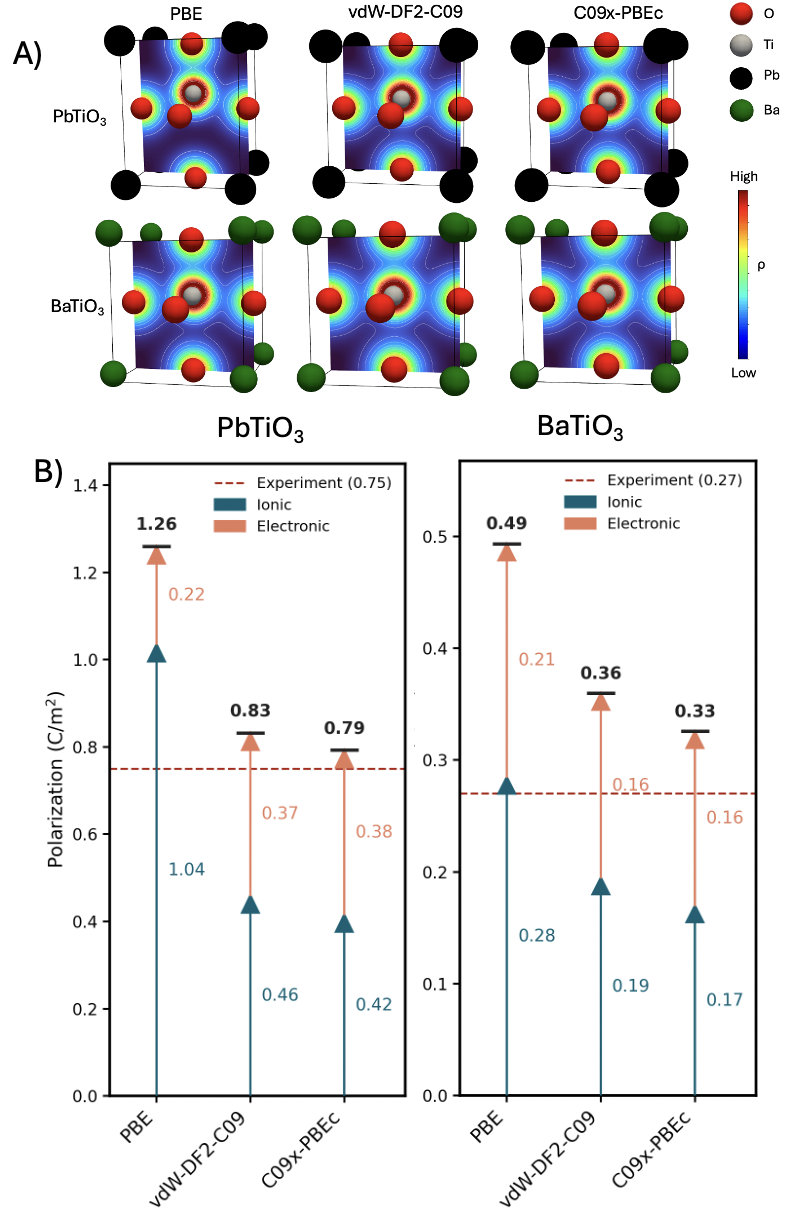} 
    \caption{A) Charge density plots for \ce{PbTiO3} (top) and \ce{BaTiO3} (bottom) with PBE, vdW-DF2-C09 and C09x-PBEc functionals B) Comparison of ionic and electronic contributions to overall polarization for with PBE, vdW-DF-C09, vdW-DF2-C09, C09x-PBEc for \ce{PbTiO3} and \ce{BaTiO3}.}
    \label{Figure2}
\end{figure}

These effects can be further investigated through analysis of the electronic structure, which, for simplicity, we only do using PBE, vdW-DF2-C09 and C09-PBEx.
%
%
Electron density plots (Fig.~\ref{Figure2}A) show that there is no overlap in electron density between the central Ti atom of the unit cell and the axial O atom of the tetragonal structure of \ce{PbTiO3} with PBE, whilst substantial overlap is seen for vdW-DF2-C09 and C09x-PBEc.
This is due to the increase in short range repulsion from the PBE exchange term which leads to overestimation of the unit cell in the $z$-direction and subsequently weaker hybridization between the O 2p orbitals and 3d of Ti. 
In contrast, the smaller short-range repulsion in the $z$-directions of C09 exchange leads to a more reasonable $c$ parameter, and to a stronger hybridization between the central Ti atom and the axial O. 
For \ce{BaTiO3}, the differences are less pronounced, where for all functionals density overlap is seen between the central Ti atom and the apical O, although to a different degree, following the same trend as PbTiO$_3$. 
%

It is important to note that the overestimation of the $c$ lattice parameter in PBE is the main reason of its overestimation of the polarization as
%
%
presented in Fig. \ref{Figure2}B. 
For \ce{PbTiO3}, the ionic component of the polarization (blue arrow) dominates compared to the electronic one in PBE, being 1.04 Cm$^{-2}$ compared to the electronic 0.22 Cm$^{-2}$. 
%
%
In comparison, for vdW-DF2-C09 and C09x-PBEc, the smaller lattice parameter $c$ leads to a reduce ionic component of P; the ionic and electronic components are more equal with ionic and electronic values of 0.42 and 0.38 Cm$^{-2}$ for vdW-DF2-C09, respectively, and 0.42 and 0.38 Cm$^{-2}$ for C09x-PBEc. 
%
%
For \ce{BaTiO3}, the ionic and electronic components of the total spontaneous polarization remain contribute roughly equally across all functionals, 
however it can be observed that, also in this case, the shorter lattice $c$ parameter obtained with C09 exchange leads to a smaller ionic polarization.
Thus we see that functionals that overestimate the $c$ lattice parameter in the tetragonal phase also overestimate the polarization.
%

%
Similar trends are seen when analysis the calculated Born effective charges (BEC) as presented in Table \ref{tab:born_split}. 
It is seen that PBE 
consistently underestimate the magnitude of the BECs (Table III), where functionals containing C09 exchange perform comparatively better. 
%

\begin{table*}[ht]
\caption{Calculated Born effective charges ($Z^*$) for \ce{PbTiO3} and \ce{BaTiO3} compared to experiment.}
\centering
\label{tab:born_split}
\small 
\setlength{\tabcolsep}{18pt} 
\begin{tabular}{llcccc}
\toprule
\textbf{System} & \textbf{Properties} & \textbf{PBE} & \textbf{vdW-DF2-C09} & \textbf{C09x-PBEc} & \textbf{Exp}. \cite{turik_origin_2000, ghosez_born_1995} \\
\midrule
\textbf{PbTiO$_3$}  & $Z^*$ (Ti)      & 4.59 & 5.35 & 5.54 & 7.60  \\
                    & $Z^*$ (Pb)      & 3.29 & 3.53 & 3.57 & 3.90  \\
                    & $Z^*$ (O$_{\bot}$)& -1.94 & -2.19 & -2.23 & -2.56   \\
                    & $Z^*$ (O$_{\parallel}$)    & -4.01 & -4.61 & -4.79 & -5.83  \\
\midrule
\textbf{BaTiO$_3$}  & $Z^*$ (Ti)      & 4.74 & 5.72 & 5.74 & 6.90  \\
                    & $Z^*$ (Ba)      & 2.90 & 2.77 & 2.82 & 2.90  \\
                    & $Z^*$ (O$_{\bot}$)& -1.84 & -1.95 & -1.97 & -2.4  \\
                    & $Z^*$ (O$_{\parallel}$)    & -4.04 & -4.77 & -4.81 & -4.8  \\
\bottomrule
\end{tabular}
\end{table*}

%
%
%
%

%
%

\begin{figure}
    \centering
    \includegraphics[width=1.0\linewidth]{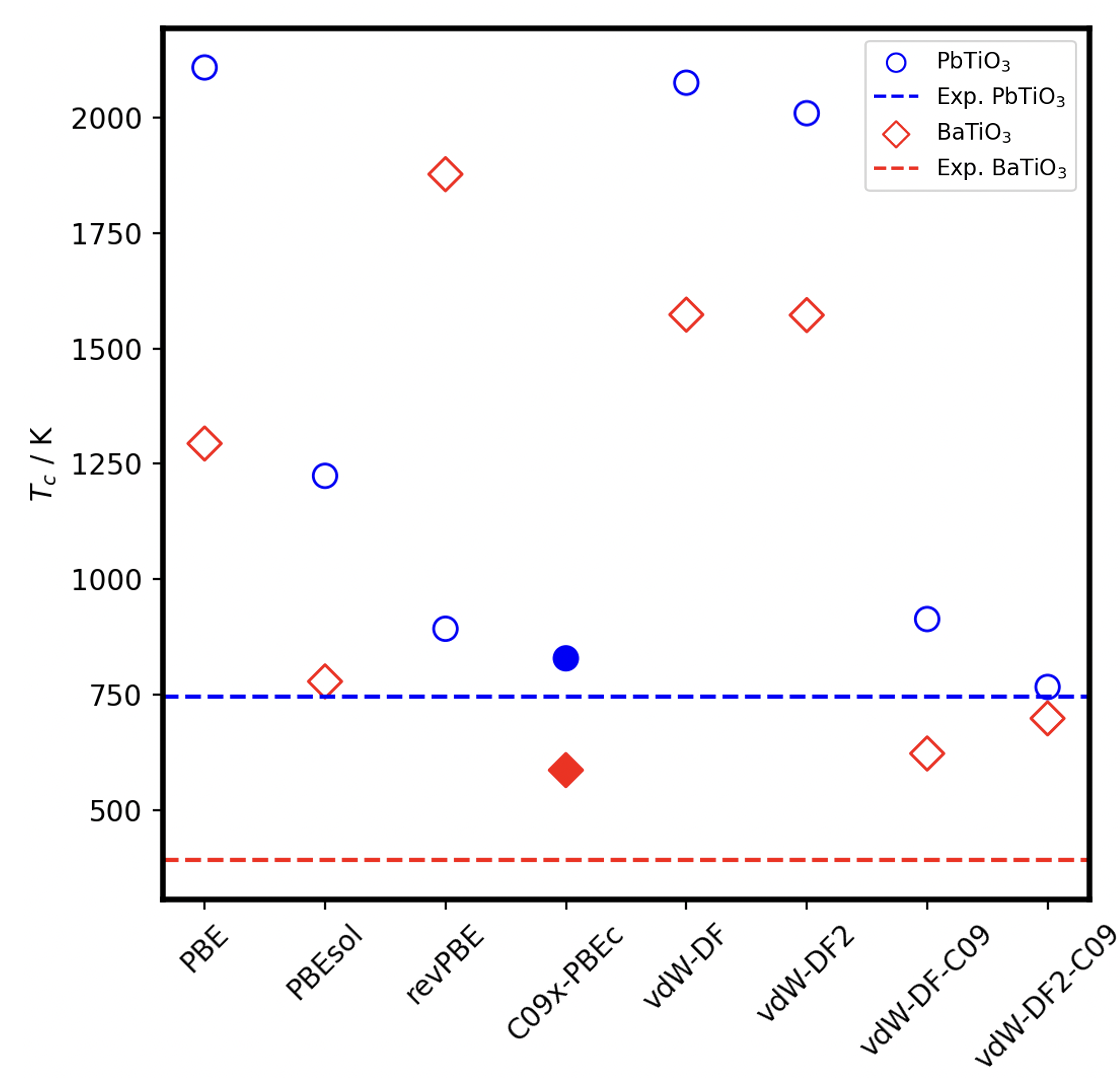} 
    \caption{Estimated $T_{c}$ for \ce{PbTiO3} (blue circles) and \ce{BaTiO3} (red diamonds) with DFT compared to experiment}.
    \label{Figure4}
\end{figure}

An accurate computed value of the ferroelectric polarization is also necessary to estimate the ferroelectric-to-paraelectric phase transition temperature, which may be estimated using Landau theory, and describes the relationship between the potential energy landscape and the polarization, via: 

\begin{equation}
\Delta E(P) = aP^{2} + bP^{4}
\end{equation}

where $\Delta E$ is the total energy relative the minimum energy of the tetragonal phase, and $P$ is the calculated polarization. 
From this it is possible to estimate the phase transition temperature, $T_{c}$, for each DFT functional via:

\begin{equation}
T_{c} = \gamma P^{2}
\end{equation}

where $\gamma$ is a constant derived from the experimental transition temperature and spontaneous polarization, and $P$ is the spontaneous polarization from calculations.
The estimated $T_{c}$ 
for \ce{PbTiO3} and \ce{BaTiO3} with each DFT functional are presented in Fig. \ref{Figure4}. 
As expected, PBE overestimates $T_{c}$ for \ce{PbTiO3} (\ce{BaTiO3}), predicting a transition temperature of $2180$ K ($1294$ K) compared to experimental value of $737$ K ($393$ K). 
Conversely, functionals containing C09 exchange perform much better (full reported values are available in the SI, Table S1). 
%
%

\begin{figure}
    \centering
    \includegraphics[width=1.0\linewidth]{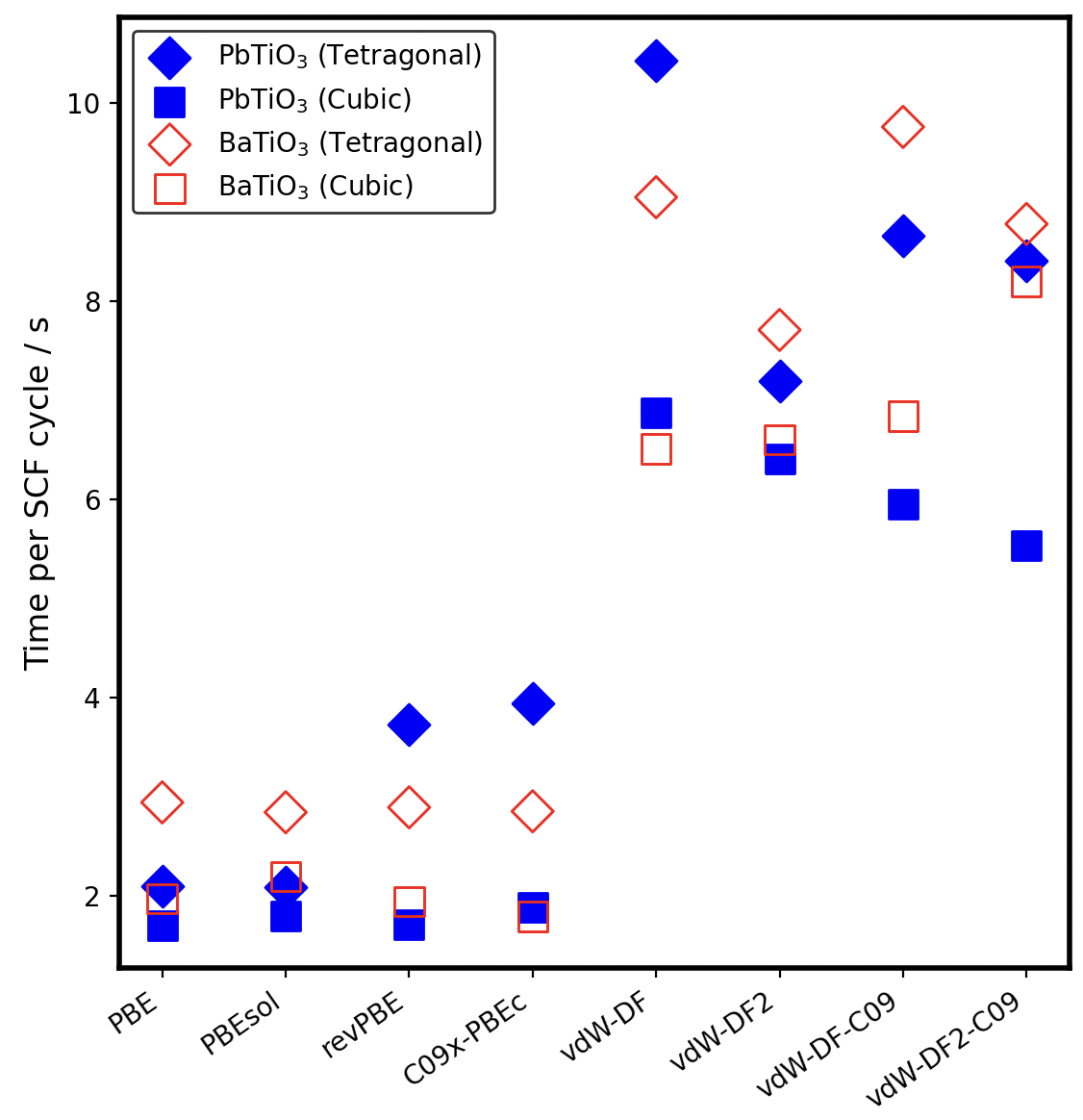} 
    \caption{Time per SCF  cycle (seconds) for energy calculations with PBE, PBEsol, C09x-PBEc, revPBE, vdW-DF, vdW-DF, vdW-DF-C09, vdW-DF2-C09 for tetragonal \ce{PbTiO3} (filled blue diamonds), cubic \ce{PbTiO3} (filled blue squares), tetragonal \ce{PbTiO3} (open red diamonds) and cubic \ce{PbTiO3} (open red squares). Time per SCF cycle was calculated by dividing wall time by number of iterations per single SCF cycle.}.
    \label{Figure3}
\end{figure}

We have thus shown that vdW-DF-C09, vdW-DF2-C09 and C09x-PBEc have a similar accuracy in describing the structural and electronic properties of ferroelectric perovskites, and that they perform better than functionals with GGA exchange.
C09x-PBEc has a further advantage over vdW-DF-C09 and vdW-DF2-C09, which is a significantly smaller computational overhead.
Fig. \ref{Figure3} shows that for \ce{PbTiO3} and \ce{BaTiO3} 
the functionals considered in this work may be separated into two groups based on the computational overhead. 
Firstly, functionals containing PBE terms, including C09x-PBEc, have the lowest overhead and secondly, those with non-local correlation have the largest overhead, with vdW-DF2-C09 being four times more expensive compared to PBE.
Instead, C09x-PBEc retains the accuracy of vdW-DF2-C09 but is only twice as expensive as PBE for tetragonal \ce{PbTiO3} and of the same order of magnitude as PBE for tetragonal \ce{BaTiO3}. 
%
%
Therefore, we emphasize that C09x-PBEc is an excellent choice in terms of speed and accuracy and would be suitable for high-throughput investigations of ferroelectric perovskite systems, in addition to being used on datasets for training MLIPs, which we demonstrate in the following section. 

\section{Applications to Machine Learning Interatomic Potentials}

We now 
%
%
compare the accuracy of an MLIP trained using data sets constructed with C09x-PBEc and foundational models trained with PBE in predicting the properties and behavior of \ce{PbTiO3}.
Full values are reported in Table \ref{mlip}.
%
%
%
The MLIP model trained on C09x-PBEc accurately captures the lattice parameters for \ce{PbTiO3} compared to DFT within $0.8$ \% and $1.5$\% difference for the $a$ and $c$ parameter of the tetragonal phase, respectively.
For the cubic phase, the trained model reproduced the $a$ parameter also within $0.5$\% difference. 
The performance of the MLIP is benchmarked against foundational MACE MLIPs models, which are trained on data from PBE+U~\cite{batatia_foundation_2025}.
MACE-MPA-0 foundational model overestimates the $c$ of the tetragonal phase by $6$\% compared to experiment. 
Interestingly, the addition of dispersion corrections to the foundational model improves the performance, where MACE-MP-0-D3 reproduces the experimental results with a $3$ \%, a significant improvement on MACE-MPA-0. \\
Overall, the results show that a model trained on C09x-PBEc accurately captures the structural parameters compared to experiment and the reference DFT.

\begin{figure}
    \centering
    \includegraphics[width=1.0\linewidth]{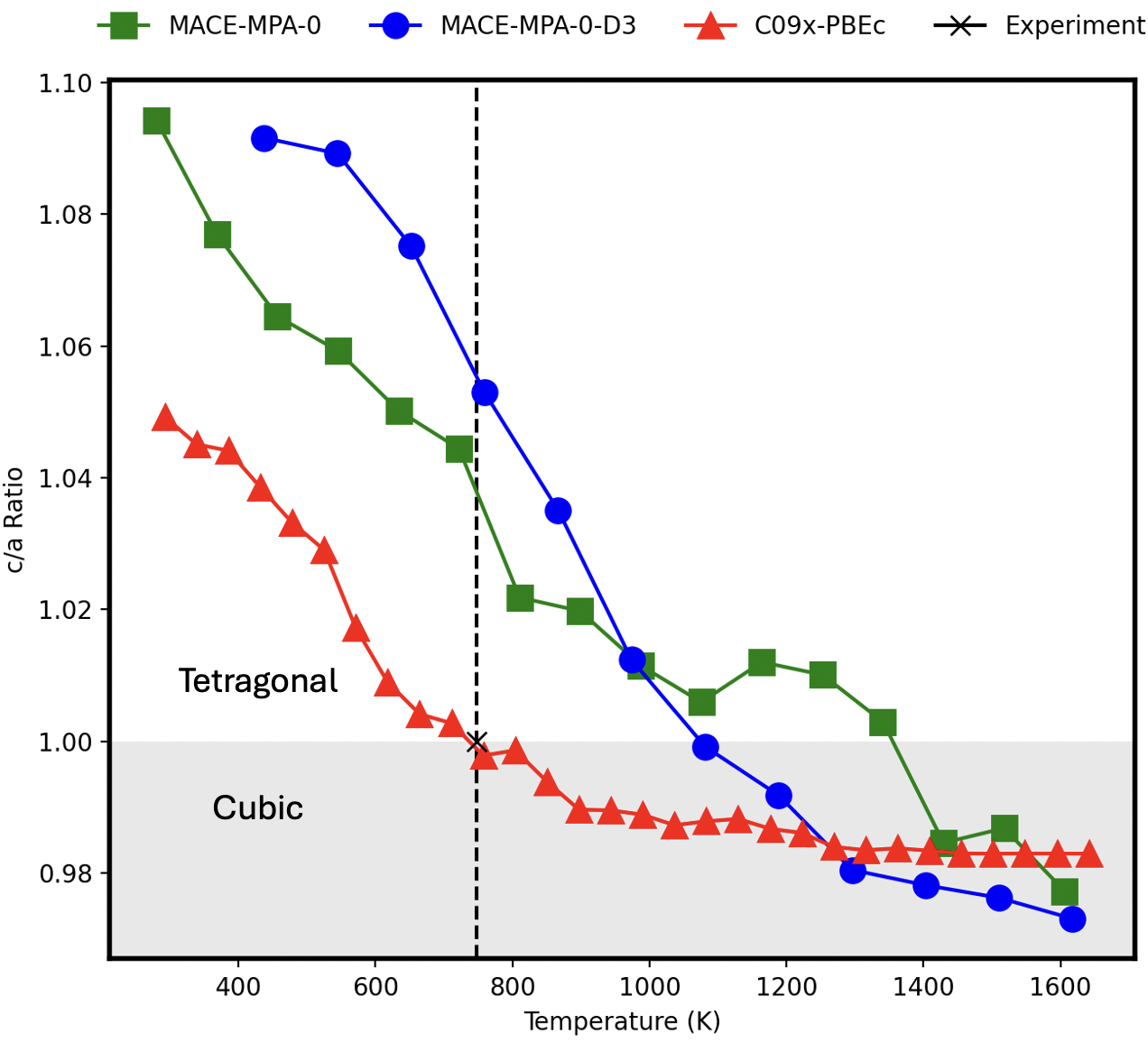} 
    \caption{$c/a$ ratio of \ce{PbTiO3} as a function of temperature calculated from NPT MD simulations using MLIPs: MACE-MPA-0 (green), MACE-MPA-0-D3 (blue), C09x-PBEc (red), with experimental transition temperature (black dashed line). }
    \label{Figure7}
\end{figure}

From the previous section, it is seen that using Landau theory, the transition temperature, $T_{c}$, from ferroelectric (tetragonal) to cubic (paraelectric) can be estimated. 
%
%
Using the trained MLIP and foundation models, $T_{c}$ can directly be calculated through NPT simulations with increasing temperature (Fig. \ref{Figure7}). 
Fig.~\ref{Figure7} shows that MACE-MPA-0, as with PBE, overestimates $T_{c}$ with a value 1429 K from this simulation. 
Again, improvements are seen with the inclusion of dispersion corrections MACE-MPA-0-D3 which has a calculated $T_{c}$ of 1082 K, but model trained on C09x-PBEc accurately calculates $T_{c}$ with a value of 738 K which is commensurate with the experimental value. 

%
Thus, the trained model accurately captures the structural properties along with $T_{c}$ from the ferroelectric to the paraelectric phase, 
demonstrating that the C09x-PBEc functional is a suitable choice for modeling phenomena pertaining to ferroelectric perovskites, and is also applicable to the training of MLIP models, which pave the way for modeling more complex phenomena where larger system sizes and time scales may be needed.

\begin{table*}[ht]
\caption{Calculated structural properties of \ce{PbTiO3} and \ce{BaTiO3} using different MACE models compared to experiment.}
\label{mlip}
\centering
\begin{tabular*}{\textwidth}{l@{\extracolsep{\fill}}lcccc}
\toprule
\textbf{Phase} & \textbf{Properties} & \textbf{MACE-MPA-0} & \textbf{MACE-MPA-0-D3} & \textbf{C09x-PBEc} & \textbf{Exp.} \\
\midrule
Cubic      & a / \AA   & 3.98   & 3.95   &     \textbf{3.91}       & 3.93 \\
Tetragonal & a / \AA   & 3.83   & 3.90   &     \textbf{3.87}       & 3.90 \\
           & c / \AA   & 4.90   & 4.26   &     \textbf{4.08}       & 4.15 \\
           & T$_c$ / K & 1429   & 1082   &     \textbf{738}      & 747  \\
\bottomrule
\end{tabular*}
\end{table*}
 
\section{Conclusions}

In conclusion, we have examined the accuracy of a selection of DFT functionals in  calculating the properties of ferroelectric perovskites and we introduce a new functional, C09x-PBEc, which is an excellent trade off between computational efficiency and structural accuracy. 

For \ce{PbTiO3} and \ce{BaTiO3}, GGA and vdW-DF functionals overestimate structural parameters and spontaneous polarization 
while functionals with C09 exchange perform significantly better. 
The improved performances is attributed to the form of the exchange enhancement factor of the exchange component of the XC functional.
In C09 exchange the EEF reduces short range repulsion compared to PBE, thus reducing the overestimation of the $c$ lattice parameter and subsequently the spontaneous polarization. 
%
%
%
The C09x-PBEc functional makes significant improvements on standard PBE, where for tetragonal \ce{PbTiO3} the calculated the $c$ parameter with a percentage error of $1.7$\% and of $5$\% for the polarization, but at a similar computational cost to PBE. 
%

Finally, we trained a MLIP for \ce{PbTiO3} with C09x-PBEc. The model accurately captures the properties of \ce{PbTiO3} compared to reference DFT data and experiment. Using the model we calculate the phase transition of \ce{PbTiO3} from tetragonal to cubic using molecular dynamics. The model trained on C09x-PBEc predicts a phase transition temperature, $T_{c}$ of $737$ K compared to $747$ K from experiment while MACE foundational models which are trained on PBE highly overestimate $T_C$. 

Our results show that the choice of theory has a significant effect on macroscopic properties of ferroelectric perovskites, and careful consideration must be given when selecting level of DFT for training MLIPs.

\section{Associated Content}

The accompanying supporting information contains further details of methods and results. All structures associated with the presented work are available on the NOMAD repository (DOI:10.17172/nomad.mdqt-vfjt). 
 
\section{Author Contribution}

The project was conceptualized by CG and OTB. Calculations and analysis were performed by OTB. Supervision was provided by CG. The article was drafted and critical revised by all authors, and all have provided approval of publication.

\section{Acknowledgments}

CG was supported by the EPSRC through a New Investigator Award (grant number UKRI132). CG and OTB acknowledge HPC support through King's Computational Research, Engineering and Technology Environment (CREATE), King's College London (2026).

\section{Conflicts of Interest}

There are no conflicts to declare.

\bibliography{references}

\end{document}